\documentclass[12pt]{spieman}  % 12pt font required by SPIE;
\usepackage{amsmath,amsfonts,amssymb}
\usepackage{graphicx}
\usepackage{setspace}
\usepackage{tocloft}
\usepackage{threeparttable}
\usepackage[subrefformat=parens]{subcaption}
\title{High-contrast H$\alpha$ imaging with Subaru/SCExAO+VAMPIRES}

\author[a,b,c]{Taichi Uyama}
\author[d,e]{Barnaby Norris}
\author[f]{Nemanja Jovanovic}
\author[g]{Julien Lozi}
\author[d]{Peter Tuthill}
\author[g,h,i]{Olivier Guyon}
\author[f]{Tomoyuki Kudo}
\author[i]{Jun Hashimoto}
\author[j,i,c]{Motohide Tamura}
\author[k]{Frantz Martinache}

\affil[a]{Infrared Processing and Analysis Center, California Institute of Technology, 1200 E. California Blvd., Pasadena, CA 91125, USA}
\affil[b]{NASA Exoplanet Science Institute, Pasadena, CA 91125, USA}
\affil[c]{National Astronomical Observatory of Japan, 2-21-1 Osawa, Mitaka, Tokyo 181-8588, Japan}
\affil[d]{Sydney Institute for Astronomy, School of Physics, Physics Road, University of Sydney, NSW 2006, Australia}
\affil[e]{AAO-USyd, School of Physics, University of Sydney 2006}
\affil[f]{Department of Astronomy, California Institute of Technology, 1200 E. California Blvd., Pasadena, CA 91125, USA}
\affil[g]{Subaru Telescope, National Astronomical Observatory of Japan, 650 North A‘ohoku Place, Hilo, HI96720, USA}
\affil[h]{Steward Observatory, University of Arizona, Tucson, AZ 85721, USA}
\affil[i]{Astrobiology Center of NINS, 2-21-1 Osawa, Mitaka, Tokyo 181-8588, Japan}
\affil[j]{Department of Astronomy, The University of Tokyo, 7-3-1, Hongo, Bunkyo-ku, Tokyo 113-0033, Japan}
\affil[k]{Universit$\acute{e}$ C$\hat{o}$te d'Azur, Observatoire de la C$\hat{o}$te d'Azur, CNRS, Laboratoire Lagrange, France}

\cftpagenumbersoff{figure}
\cftpagenumbersoff{table} 
\begin{document} 
\maketitle

\begin{abstract}
%\textcolor{red}{TU: preparing for submitting the manuscript to arxiv - revised/added lines in bold type are stored in the backup}
We present current status of H$\alpha$ high-contrast imaging observations with Subaru/SCExAO+VAMPIRES. Our adaptive optics correction at optical wavelengths in combination with (double) spectral differential imaging (SDI) and angular differential imaging (ADI) was capable of detecting a ring-like feature around omi Cet and the H$\alpha$ counterpart of jet around RY Tau.
We tested the post-processing by changing the order of ADI and SDI and both of the contrast limits achieved $\sim10^{-3}-5\times10^{-4}$ at $0.3^{\prime\prime}$, which is comparable to other H$\alpha$ high-contrast imaging instruments in the southern hemisphere such as VLT/SPHERE, VLT/MUSE, and MagAO.
Subaru/VAMPIRES provides great opportunities for H$\alpha$ high-contrast imaging for northern hemisphere targets.

\end{abstract}

% Include a list of up to six keywords after the abstract

% Include email contact information for corresponding author
{\noindent \footnotesize\textbf{*}Taichi Uyama,  \linkable{tuyama@ipac.caltech.edu} }

\begin{spacing}{2}   % use double spacing for rest of manuscript

\section{Introduction}
\label{sec: Introduction}

%uniqueness of H$\alpha$ observation (not only emission but also absorption) - accretion mechanism, shock, jet, atmosphere, etc... (example?)
%Combination of AO - high-contrast imaging, useful to detect H$\alpha$ near a star while avoiding stellar halo

Observing hydrogen lines provides fruitful opportunities to investigate mass accretion, shocks, jets, atmospheres, and other astrophysical phenomena.
A variety of observations have been implemented but previous observations basically targeted an isolated object or multi-objects that can be spatially resolved under seeing-limited conditions.
An improvement in instrumentation has enabled better angular resolution than set by the seeing limit - adaptive optics (AO) \cite{Beckers1993} systems can make real-time corrections to wavefront distortions by the Earth's atmosphere using a guide star and delivers a point spread function (PSF) close to the diffraction-limited one.
In particular, AO is very important for high-contrast imaging of exoplanets or protoplanetary disks by removing stellar halo and instrumental speckles that bury such faint signals in the stellar halo.
However, the AO correction at optical wavelengths was difficult in the early years of the AO instruments because the Fried parameter ($\propto \lambda^{6/5}$) \cite{Fried1965} is so small at optical wavelengths and hence a large number of actuators are required across the telescope aperture as well as fast temporal operation for wavefront sensing.
Therefore, classical AO instruments were mostly applied near-infrared (NIR) wavelengths (e.g. $JHKLM$ band), which benefit exoplanetary science because low-mass objects have less contrast with respect to their host stars in the NIR and thus high-contrast imaging at these wavelengths provides the best sensitivity.

Recently, further improvements in instrumentation have made it possible to operate AO correction at optical wavelengths (e.g. VLT/MUSE \cite{Bacon2010}, VLT/SPHERE \cite{Beuzit2019}, MagAO \cite{Males2014}, and Subaru/SCExAO \cite{Lozi2018}) and have kicked off a new era of high-contrast imaging at H$\alpha$ ($\lambda=$656.28 nm).
One of the most important subjects of H$\alpha$ observation with AO is active mass accretion onto protoplanets (e.g. PDS 70 bc \cite{Wagner2018, Haffert2019}). 
Recent disk observations with ALMA or NIR polarimetric observations have shown a variety of asymmetric features within $1^{\prime\prime}$ that may be related to planet formation (e.g. gap, ring, or spiral \cite{Andrews2018, Dong2018}), but the number of confirmed protoplanets is still smaller than the number of predictions of potential protoplanets. Therefore, the planet formation mechanism is still controversial and exploring for planets at such inner regions while avoiding the stellar halo with high-performance AO correction is important.
The performance at optical wavelengths with current instruments highly depends on airmass. For example, the performance of MUSE/NFM (Narrow-Field-Mode) empirically decreases with airmass $\geq1.6$ \footnote{MUSE manual: \url{https://www.eso.org/sci/facilities/paranal/instruments/muse/doc.html}}. We also checked some archival MUSE data \footnote{\url{http://archive.eso.org}} and compared airmass and full width at half maximum (FWHM) in the H$\alpha$ channel (see Table \ref{tab: MUSE log}), which confirms the degradation of AO correction at high airmass.
Poor AO correction leads to ineffective post-processing to remove the stellar halo and to achieve high contrast.
As VLT and MagAO are located in the southern hemisphere, their observations of northern targets, e.g. the Taurus star forming region, may not have sufficient sensitivity to detect H$\alpha$ in the vicinity of a star.

Here we have installed a new observing mode of The Visible Aperture Masking Polarimetric Imager for Resolved Exoplanetary Structures (VAMPIRES) at Subaru Telescope~\cite{Norris2015}, which is the only instrument capable of H$\alpha$ imaging fed by AO in the northern hemisphere at the moment.
In this paper we present the current performance of high-contrast H$\alpha$ observations with VAMPIRES. 
The effective bandwidth of H$\alpha$ is narrower than the widths of broad-band filters used in the optical-NIR astronomy \cite{Tokunaga2002,Doi2010} and thus investigating such emissions or absorption's requires a specified narrow-band filter or spectroscopic instrument with $R\gtrsim1000$.
VAMPIRES adopts a narrow-band filter and details of the specifications are described in Section \ref{sec: VAMPIRES Specifications}.
Section \ref{sec: Science Verification} presents our engineering observations and the results.
Finally we summarize our work and briefly mention future prospects with AO upgrades in Section \ref{sec: Summary and Fugure Prospects}.

\begin{table}
\caption{VLT/MUSE archival data of YSOs} 
\label{tab: MUSE log}
\centering
\begin{threeparttable}[h]
\begin{tabular}{|c|c|c|c|c|c|c|} 
\hline
\rule[-1ex]{0pt}{3.5ex} Target & Association & Date [UT]  & Airmass & $R$ mag\tnote{1} & $H$ mag\tnote{2}  & FWHM [mas]\\ \hline\hline
PDS 70 & Centaurus & 2018 June 20 & 1.05 & 11.6 & 8.82  & $\sim$60 \\ \hline
CIDA-9 & Taurus & 2019 Nov 2 & 1.56 & 15.6 & 11.9 & $\sim$80 \\ \hline %SIMBAD: IRAS 05022+2527
CI Tau & Taurus & 2019 Nov 6 & 1.60 & 12.2 & 8.43 & $\sim$130 \\ \hline
GO Tau & Taurus & 2019 Nov 3 & 1.70 & 14.2 & 9.78 & $\sim$150 \\ \hline
DS Tau & Taurus & 2019 Nov 2 & 1.74 & 11.8 & 8.60 & $\sim$170 \\ \hline
\end{tabular}
\begin{tablenotes}
\item[1]{UCAC4 catalogue \cite{Zacharias2012-UCAC4}}
\item[2]{2MASS \cite{Cutri2003-2MASS}}
\end{tablenotes}
\end{threeparttable}
\end{table}

\section{SCExAO+VAMPIRES Specifications} \label{sec: VAMPIRES Specifications}

The VAMPIRES instrument\cite{Norris2015} is a module of the Subaru Coronagraphic Extreme Adaptive Optics (SCExAO) instrument at the Subaru telescope\cite{Jovanovic2015,Lozi2018,Currie2019}. SCExAO, installed behind Subaru's facility adaptive optics AO188, performs a second stage of wavefront correction using a 2000-element deformable mirror and a visible pyramid wavefront sensor (PyWFS). The control loop typically corrects the wavefront at a frequency of 2000~Hz. The near-infrared light (0.95 to 2.4~$\mu$m) typically goes through a coronagraph and is recorded using the integral field spectrograph CHARIS. In median seeing conditions, SCExAO provides extreme-AO performance with NIR Strehl ratios over 80\%. In visible light (600 to 900~nm), the light not used by the PyWFS is sent towards the VAMPIRES module. Optical Strehl ratio's with SCExAO depend on conditions (e.g. seeing, airmass, and brightness of a guide star) and we show examples of estimated Strehl ratios in our observations in Section \ref{sec: Observation}.

VAMPIRES is capable of performing diffraction limited, polarimetric imaging at visible wavelengths (600 to 800~nm). It uses two electron-multiplying CCD (EMCCD) cameras which in normal operation each record an orthogonal polarisation, which in combination with a ferroelectric liquid crystal modulator and half-waveplate allow precise polarimetric differential imaging (PDI). The high speed of the cameras (up to $\sim$1000 frames/sec depending on subwindow size) allows lucky imaging techniques to be used to enhance resolution, and also include non-redundant masks for super-diffraction limited imaging. 

\begin{figure}
    \centering
    \includegraphics[width=\textwidth]{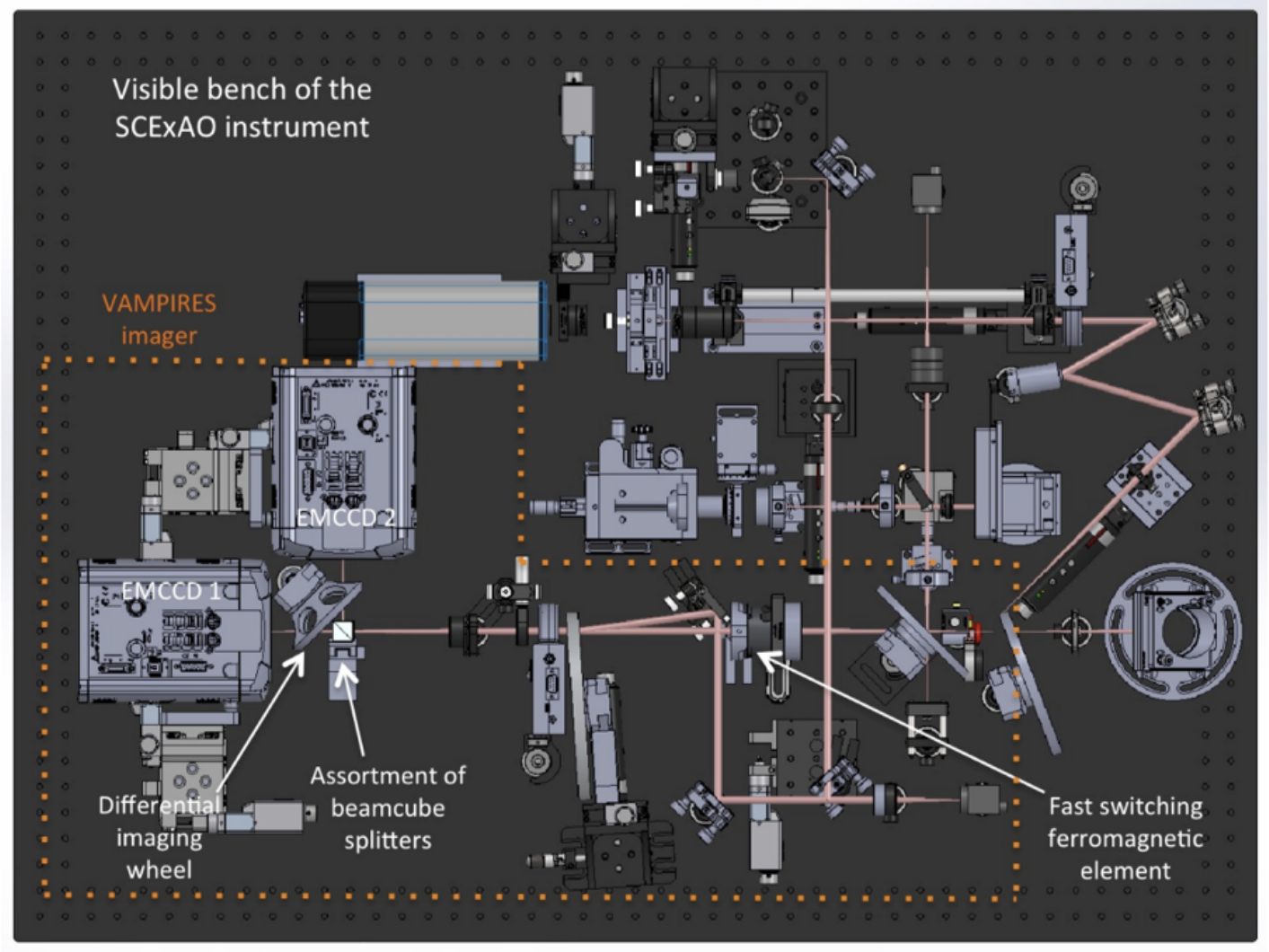}
    \caption[Schematic view of SCExAO+VAMPIRES]{Schematic view of the entire SCExAO visible bench. The VAMPIRES instrument is highlighted by the dashed orange box.}
    \label{fig: SCExAO schematic}
\end{figure}

With the upgrade presented here, the polarisation-splitting optics can be automatically interchanged for wavelength-splitting optics, recording simultaneous images in a narrow band centred at H$\alpha$ ($\lambda_{\rm cen}=$656.3 nm, $\Delta\lambda=1.0$ nm) in one camera and an adjacent continuum bandpass ($\lambda_{\rm cen}=$647.68 nm, $\Delta\lambda=2.05$ nm) in the other. The filter information is summarized in Appendix \ref{sec: Filter Information}.
The current field of view (FoV) of VAMPIRES is $\sim3.3^{\prime\prime}\times3.3^{\prime\prime}$.
Moreover, the choice of which of these filters is positioned in front of which camera can be rapidly switched during observation, to allow non-common path errors to be mitigated by a `double-difference' approach during data reduction (see Figure \ref{fig: SCExAO schematic} and the instrument papers for details \cite{Norris2015,Lozi2018}). 
The splitting of the light before encountering the filters is performed via a non-polarizing beamsplitter, to minimise mixing of polarization effects with spectral differential imaging (SDI; see Section \ref{sec: Science Verification}) \cite{Smith1987} signal.

\section{Science Verification} \label{sec: Science Verification}
\subsection{Target Selection}
For science verification we target omicron Ceti (omi Cet) and RY Tau. These targets are reported to have H$\alpha$ emissions.
RY Tau is an intermediate-mass classical T Tauri Star (TTS) in the Taurus star forming region (mass: $\sim1.9M_\cdot$, age: 4.5 Myr \cite{Garufi2019}). This TTS is known to eject a jet \cite{Cabrit1990, Agra-Amboage2009} and the H$\alpha$ counterpart was imaged by SPHERE/ZIMPOL \cite{Garufi2019}. omi Cet is a Mira variable star, undergoing massive stellar pulsations wherein shocks are expected to produce H$\alpha$ emission\cite{Gillet1983}. 
While detectable in spectroscopy, this is very difficult to resolve spatially due to the expected low spatial separation (several 10s of milliarcseconds) between the star and shock region.

We also use SAO 105500 ($\gamma$ Sge) data for a comparison of the post-processing results with those of omi Cet and RY Tau.
Previous spectroscopic observation proves that this single star does not present any H$\alpha$ emission/absorption features \cite{Alekseeva1997,Wittkowski2006} and is good for the comparison of the results.

\subsection{Observation} \label{sec: Observation}
SCExAO is operated under angular differential imaging (ADI) mode \cite{Marois2006} by fixing the pupil, with which we can utilize a combination of SDI and the ADI technique to detect faint objects around the target star. 
VAMPIRES uses a beam-switcher where both filters can be converted frequently and hence the spectral content of the beams is switched between the two detectors. In order to reduce further bias, which arises from non-common path aberrations and systematic differences between the detectors, and to achieve better sensitivity we utilize double differential imaging (DDI) \cite{Hinkley2009} techniques for SDI reduction (see also Figures \ref{fig: flowchart} and \ref{fig: schematic SDI} for the schematic of the post-processing).
To calibrate the plate scale, we imaged astrometric binaries HD 117902 and HIP 17954 and measured plate scale to be 6.475$\pm$0.09 mas/pix.

We observed SAO 105500, omi Cet, and RY Tau in our engineering runs on 2019 May 22, 2019 September 8, and 2020 January 31 UT respectively.
Table \ref{tab: obs log} summarizes our observations used in this study and Table \ref{tab: Strehl} summarizes the Strehl ratios of our data.
We used the beginning part of these data sets for measuring them.
To estimate the Strehl ratio we 1) computed the ratio between the total flux inside the core of the PSF ($2.44\lambda/D$ in diameter) and the total flux of the PSF inside a circle of $85\lambda/D$ (to reject the diffraction pattern created by the quilting mode of the deformable mirror) and 2) computed the same ratio calculated for a simulated perfect PSF using the shape and orientation of the pupil. The Strehl ratio is then the flux ratio of the on-sky image divided by the flux ratio of the simulated image.
The Strehl ratios measured from the single exposures indicate the performance of the 'fast-AO correction' with SCExAO+VAMPIRES. For comparison we also show the Strehl ratios with the long exposure (simply cube-combined PSF without image registration).
We note that at the data reduction stage we do image registration of each slice and then conduct the post-processing technique (see Section \ref{sec: Data Reduction}), the performance of which is indicated in the last column of Table \ref{tab: Strehl}.
For the SAO 105500 and RY Tau data we see the same characteristics in that the top 5\% PSF has the highest Strehl ratio. The single exposure of RY Tau has lower SNR because this target is faint in the optical wavelength, which may affect the background evaluation and the Strehl ratio measurement.
For the omi Cet data the long-exposure PSF has the highest Strehl ratio but this is likely related to measurement errors of the Strehl ratio and the background: we combined 3201 slices to make the long-exposure image (and $\sim$2880 slices to make the combined image) and the background values are much better estimated than the single exposure images.

Figure \ref{fig: schematic SDI} illustrates difference between State 1 (cam1: continuum, cam2: H$\alpha$) and State 2 (cam1: H$\alpha$, cam2: continuum) so that we can conduct the DDI technique to reduce effects of the non-common path aberrations, which is explained in Section \ref{sec: SDI and SDI+ADI}.
The VAMPIRES output constitutes a data cube (x, y, and time) and exposure time/cube format information is summarized in Table \ref{tab: obs log}.
Note that we replaced a H$\alpha$ narrow-band filter before the RY Tau observation and we observed SAO 105500 and omi Cet with a different H$\alpha$ filter ($\lambda=656.4$ nm, $\Delta \lambda=2.0$ nm).

\begin{table}
\caption{Observing logs} 
\label{tab: obs log}
\centering
\small
\begin{threeparttable}
\begin{tabular}{|c|c|c|c|c|c|c|c|} %% this creates two columns
%% |l|l| to left justify each column entry
%% |c|c| to center each column entry
%% use of \rule[]{}{} below opens up each row
\hline
\rule[-1ex]{0pt}{3.5ex}  
%\rule{0pt}{4ex} 
Target & Date [UT] & seeing [$^{\prime\prime}$] \tnote{1} & Airmass & $R$ mag  &\multicolumn{2}{|c|}{$t_{\rm total}$ [sec]} & remarks \\ \cline{6-7}
\rule[-1ex]{0pt}{3.5ex} & & & & & State 1 & State2 & \\ \hline\hline
\rule[-1ex]{0pt}{3.5ex} SAO 105500 & 2019 May 22 & 0.62  & 1.07  & 2.23 \tnote{2} & 256.16 \tnote{3} & 256.16 \tnote{3} &  no ADI \\
\hline 
\rule[-1ex]{0pt}{3.5ex} omi Cet & 2019 Sep 8 & 0.70 & 1.15 & 4.34 \tnote{4} & 463.96 \tnote{5} & 448.14 \tnote{6}  & smaller FoV \tnote{7}\\
\hline
\rule[-1ex]{0pt}{3.5ex} RY Tau & 2020 Jan 31 & 0.66 & 1.01 & 9.05 \tnote{4} & 2497 \tnote{8} & \dots & State 1 only, no DDI  \\
\hline 
\end{tabular}
\begin{tablenotes}
\item[1]{Mean DIMM seeing at the summit of Mauna Kea.}
\item[2]{Johnson $R$-band photometry \cite{Huang2015}.}
\item[3]{20 msec/slice $\times$ 1601 slices/cube $\times$ 8 cubes.}
\item[4]{These targets are variable at optical wavelength and we adopt photometric data provided by \href{https://www.aavso.org/}{AAVSO} on the nearest date to our observations.}
\item[5]{20 msec/slice $\times$ 3201 slices/cube $\times$ 7 cubes + 20 msec/slice $\times$ 791 slices/cube $\times$ 1 cube.}
\item[6]{20 msec/slice $\times$ 3201 slices/cube $\times$ 7 cubes.}
\item[7]{$\sim1.65^{\prime\prime}\times1.65^{\prime\prime}$.}
\item[8]{1 sec/slice $\times$ 101 slices/cube $\times$ 24 cubes + 1 sec/slice $\times$ 73 slices/cube $\times$ 1 cube.}
\end{tablenotes}
\end{threeparttable}
\end{table} 
%The Strehl ratios are estimated from long-exposure continuum images combined after dark subtraction and image registration of the selected 90\%-good data sets.

\begin{table}
\caption{Strehl Ratios} 
\label{tab: Strehl}
\centering
\small
\begin{threeparttable}
\begin{tabular}{|c|c|c|c|c|c|c|c|} %% this creates two columns
%% |l|l| to left justify each column entry
%% |c|c| to center each column entry
%% use of \rule[]{}{} below opens up each row
\hline
\rule[-1ex]{0pt}{3.5ex}  
%\rule{0pt}{4ex} 
Target & \multicolumn{4}{|c|}{Short exposure \tnote{1}} &  \multicolumn{2}{|c|}{Long exposure \tnote{2}} & Shift \& combined \tnote{3} \\ \cline{2-7}
 & $t_{\rm eq}$ [sec] \tnote{4} & Top 5\% peak & Top 30\% peak & Top 50\% peak & $t_{\rm eq}$ [sec] \tnote{4}  & &  \\ \hline\hline
\rule[-1ex]{0pt}{3.5ex} SAO 105500 & 0.02 & $\sim$44\% & $\sim$40\% &$\sim$35\% & 32.02 & $\sim$35\% & $\sim$39\%  \\ \hline 
\rule[-1ex]{0pt}{3.5ex} omi Cet & 0.02 & $\sim$41\% & $\sim$39\% & $\sim$37\% & 64.02  & $\sim$45\% & $\sim$44\% \\
\hline
\rule[-1ex]{0pt}{3.5ex} RY Tau & 1 & $\sim$9\% & $\sim$8\% & $\sim$6\% & 101 & $\sim$8\% & $\sim$9\% \\
\hline 
\end{tabular}
\begin{tablenotes}
\item[1]{We selected the single exposures whose PSFs have top 5\%, 30\%, and 50\% peaks among all the single exposures using the fitted PSF information and then estimated the Strehls from these selected PSFs. This value indicates the performance of the short-exposure AO correction.}
\item[2]{We combined a data cube into an image without image registration. This value indicates the performance of the long-exposure AO correction  (exposure time corresponds to the product of the single exposure time and the number of slices in the data cube, see also Table \ref{tab: obs log}).}
\item[3]{We selected a set of slices in a data cube (top 90\%  peaks), and then shifted all the images to align the center of the PSF, and finally combined them to make the combined image. This data set is basically used for the post-processing in this study.}
\item[4]{Equivalent integration time.}
\end{tablenotes}
\end{threeparttable}
\end{table} 

\subsection{Data Reduction} \label{sec: Data Reduction} 
As SCExAO+VAMPIRES enables fast-AO correction, we can obtain images with very short exposures. Then we can select `good' PSFs among all the data set like lucky imaging.
After dark subtraction we read all continuum slices, which do not basically include any asymmetric features related to H$\alpha$ from the central star, to fit PSFs for good-data selection and image registration. 
We used a criterion of fitted peak and selected 90\% good data sets that would then be reduced by post-processing. The typical FWHMs of the selected data sets were 45 mas (7 pix) and 55 mas (8.5 pix) for omi Cet and RY Tau, respectively.
Hereafter, we show several methods of reducing the data by combining the ADI and SDI techniques to look for differences in the outputs because, for instance, a previous VLT/SINFONI observation \cite{Christiaens2019} suggested a specific order to utilize ADI and SDI reduction techniques may change the overall sensitivity.

\subsubsection{SDI and SDI+ADI} \label{sec: SDI and SDI+ADI}
The advantage of the SDI reduction is that in principle we can subtract continuum components as a reference PSF from the H$\alpha$ image that includes both H$\alpha$ and continuum information. 
AO correction works with almost the same efficiency in both filter bands and enables us to simply subtract the continuum image from the H$\alpha$ image after correcting throughputs between cameras and filter transmission functions. The coefficients used for the throughput correction are assessed by comparing photometric results of instrumental laser PSFs.

\begin{figure}
    \centering
    \includegraphics[width=0.95\textwidth]{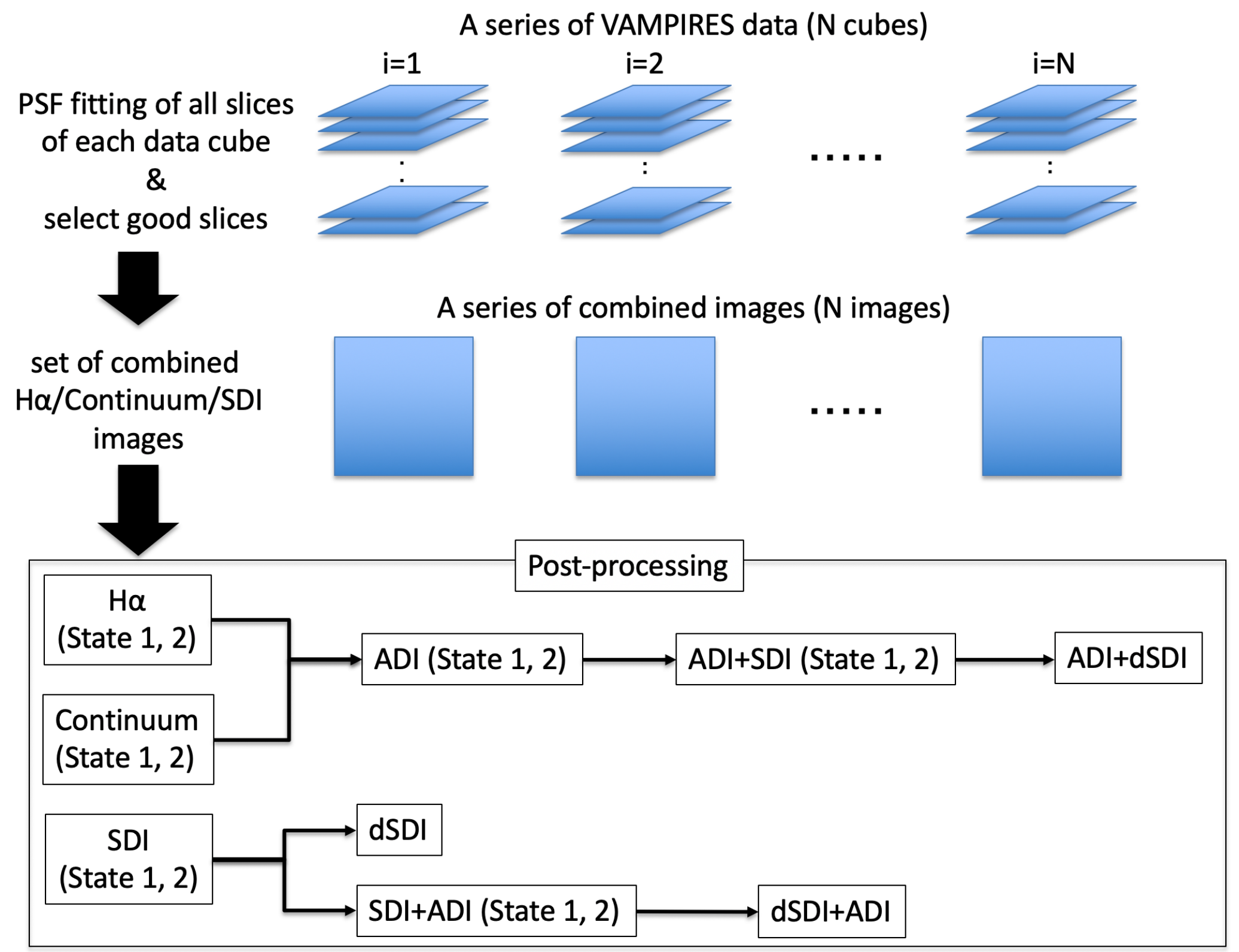}
    \caption[Flowchart of data reduction]{Flowchart of data reduction from VAMPIRES data cube through final outputs. Keyword 'dSDI' corresponds to double-SDI.}
    \label{fig: flowchart}
\end{figure}

\begin{figure}
    \centering
    \includegraphics[width=0.9\textwidth]{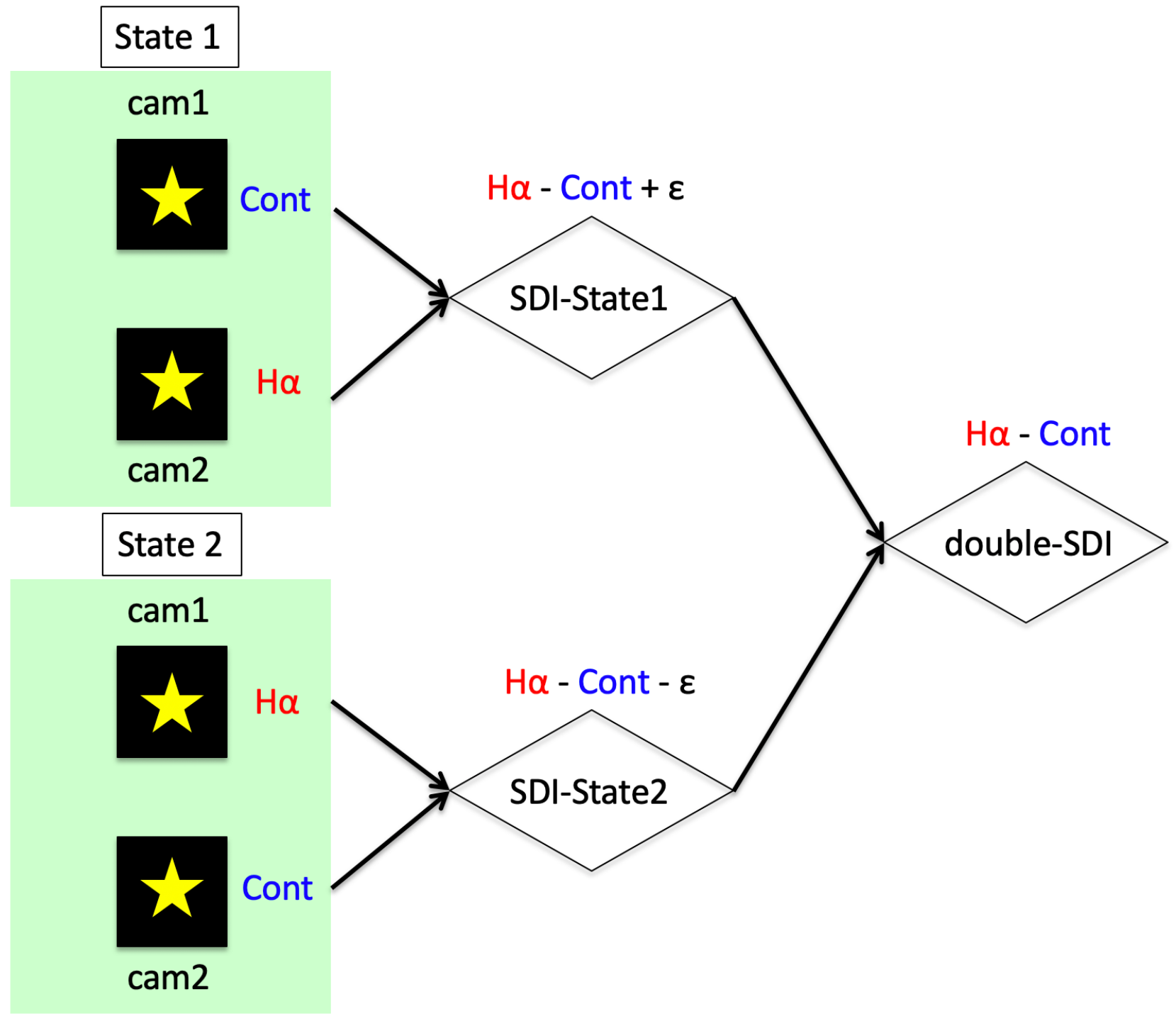}
    \caption[Schematic of double-SDI reduction]{Schematic of double-SDI with VAMPIRES H$\alpha$ mode. `H$\alpha$', `Cont', and `$\epsilon$' correspond to H$\alpha$, continuum flux, and bias between cam2 and cam1 respectively. In the State 1 we subtract the cam1 (continuum) image from the cam2 (H$\alpha$) image while in the State 2 we do the other way around.}
    \label{fig: schematic SDI}
\end{figure}

Figure \ref{fig: flowchart} illustrates a flowchart of data reduction and Figure \ref{fig: schematic SDI} shows a brief schematic of the double-SDI reduction with VAMPIRES respectively. 
VAMPIRES data consists of two states where the H$\alpha$ and continuum filters are switched with each other.
Subtracting an image taken at one of the detectors from the other image leaves non-common path aberrations, which corresponds to the bias ($\epsilon$) in Figure \ref{fig: schematic SDI}. Applying the DDI technique to the SDI reduction can further suppress the effects of the bias on the final-reduced image.
After PSF fitting of continuum slices using the Moffat function \cite{Moffat1969} we repeat producing a SDI-reduced slice by subtracting a good-continuum slice in the VAMPIRES data cube from a corresponding H$\alpha$ slice, which can attenuate the influence of the atmospheric turbulence at each short exposure. Here we used the fitted peak of each PSF as a criterion for judging good PSFs.
\\
{\bf SDI reduction}\\
Next we made a combined SDI-reduced image per data cube. As SCExAO is operated in ADI mode we derotate images by differences of parallactic angle and then combine this data set into a SDI-reduced image. At each state we conduct SDI reduction and finally obtain double-SDI-reduced image using two SDI-reduced images.
We note that in this study we mainly intend to reduce the continuum component from the H$\alpha$ image and that we do not scale the reference (continuum) image by the difference of wavelength from H$\alpha$ to attenuate the speckle noise.
\\
{\bf SDI+ADI reduction}\\
By applying the ADI technique to the residual of the SDI result we can further suppress the speckle noise.
The data set of combined SDI-reduced image per data cube can also be applied to ADI reduction (see Section \ref{sec: ADI+SDI}), which leads to two SDI+ADI images in State 1 and 2. We then conduct double-differential imaging to obtain the final double-SDI+ADI image.

\subsubsection{ADI+SDI} \label{sec: ADI+SDI}
ADI makes a likely reference PSF that includes the starlight and instrumental speckles by rotating the FoV, then subtracting it from the raw images, and finally derotating and combining the subtracted images. This post-processing technique has been widely used for high contrast imaging, and has been used to detect faint companions and/or other asymmetric features within a few arcseconds.
We made a set of images by combining selected good slices among one data cube, which is then input into ADI reduction algorithms.
In this paper we utilized Karhunen-Lo$\acute{e}$ve Image Projection algorithms (KLIP) \cite{Soummer2012-KLIP} with pyKLIP algorithms \cite{Wang2015-pyKLIP} to produce the most likely reference PSF from the set of combined SDI-reduced images, where we adopted optimization and subtraction area as large as the whole VAMPIRES FoV.

After PSF fitting of continuum slices we made both combined continuum and H$\alpha$ images per data cube.
Then we applied pyKLIP to 4 data sets (registered images of continuum/H$\alpha$ in State 1/2). Each reduced image was then used for further double-SDI reduction (see Figure \ref{fig: schematic SDI}) and we finally got the ADI+double-SDI image.
We note that the ADI+SDI reduction substantially ignores the advantage of simultaneous acquisition of H$\alpha$ and continuum at each exposure.

\subsection{Results} \label{sec: Results}
We present a variety of results that were reduced via SDI and ADI reduction techniques. 
We note that in this study we do not analyze H$\alpha$ intensities of each detection and discuss mechanisms of (possible) H$\alpha$ emissions from omi Cet, omi Cet B, and RY Tau.

\subsubsection{Omi Cet} \label{sec: Omi Cet}

\begin{figure}
\begin{tabular}{cc}
\begin{minipage}{0.5\hsize}
\centering
\includegraphics[width=0.95\textwidth]{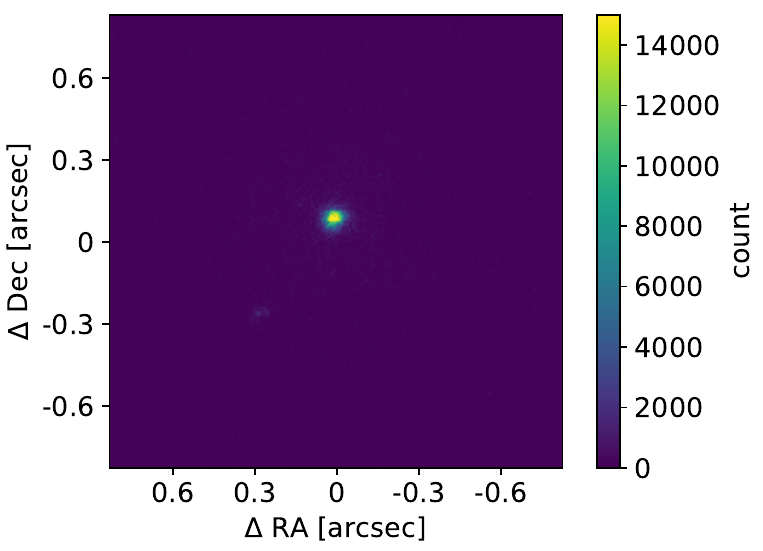}
\subcaption{Single H$\alpha$ image of omi Cet (exposure time = 20 msec) before the dark subtraction.}
\label{fig: omi cet raw}
\end{minipage}
\begin{minipage}{0.5\hsize}
\centering
\includegraphics[width=0.95\textwidth]{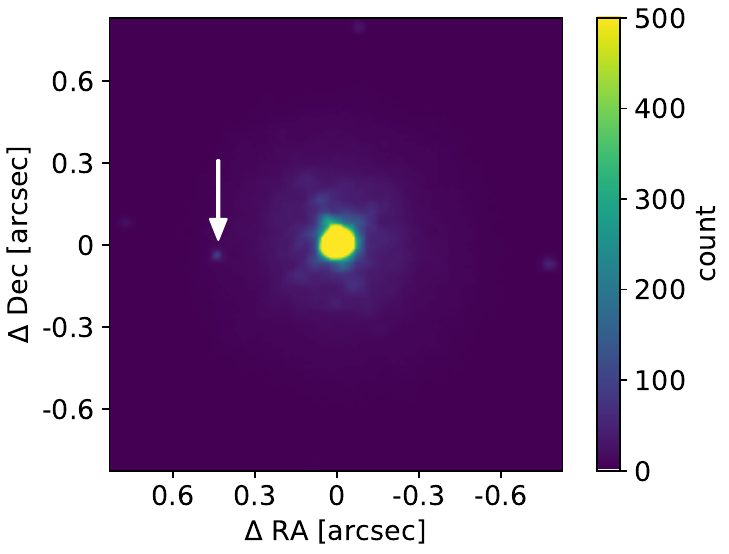}
\subcaption{Combined continuum image of omi Cet from a single data cube at State 1. The location of omi Cet B is indicated by a white arrow.}
\label{fig: omi cet comb_cont}
\end{minipage}\\
\begin{minipage}{0.5\hsize}
\centering
\includegraphics[width=0.95\textwidth]{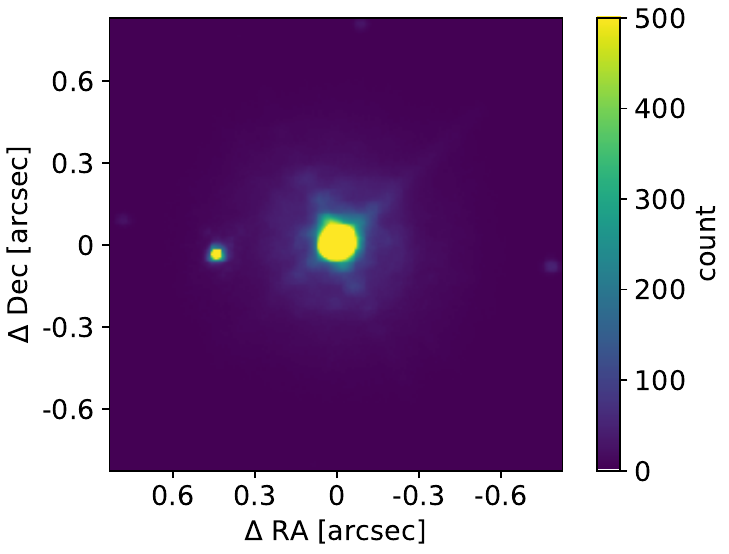}
\subcaption{Combined H$\alpha$ image of omi Cet from a single data cube at State 1. The companion (omi Cet B) is located east.}
\label{fig: omi cet comb_Ha}
\end{minipage}
\begin{minipage}{0.5\hsize}
\centering
\includegraphics[width=0.95\textwidth]{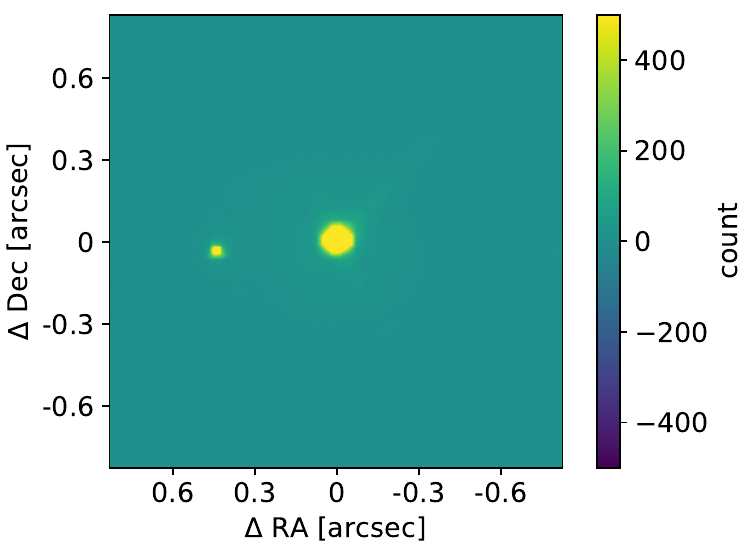}
\subcaption{SDI-reduced image (H$\alpha$-continuum) of State 1. }
\label{fig: omi cet SDI-state1}
\end{minipage}\\
\begin{minipage}{0.5\hsize}
\centering
\includegraphics[width=0.95\textwidth]{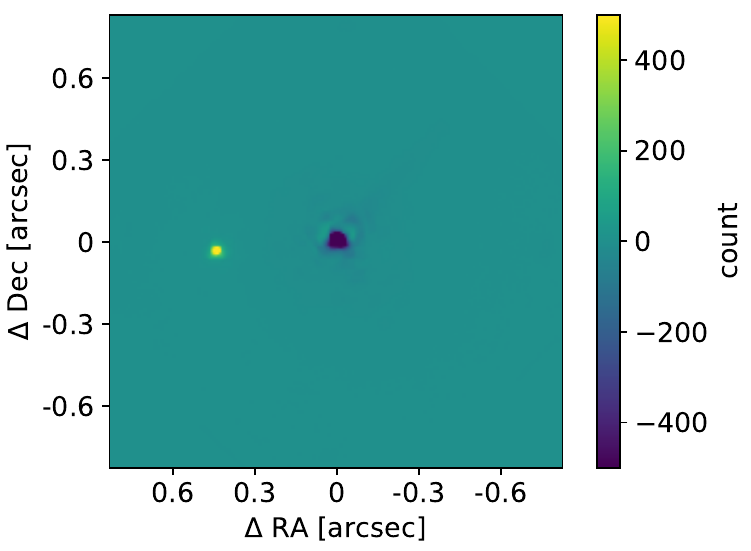}
\subcaption{SDI-reduced image (H$\alpha$-continuum) of State 2. The bias made the apparent result different from the SDI result of State 1 (see Figure \ref{fig: omi cet SDI-state1}).}
\label{fig: omi cet SDI-state2}
\end{minipage}
\begin{minipage}{0.5\hsize}
\centering
\includegraphics[width=0.95\textwidth]{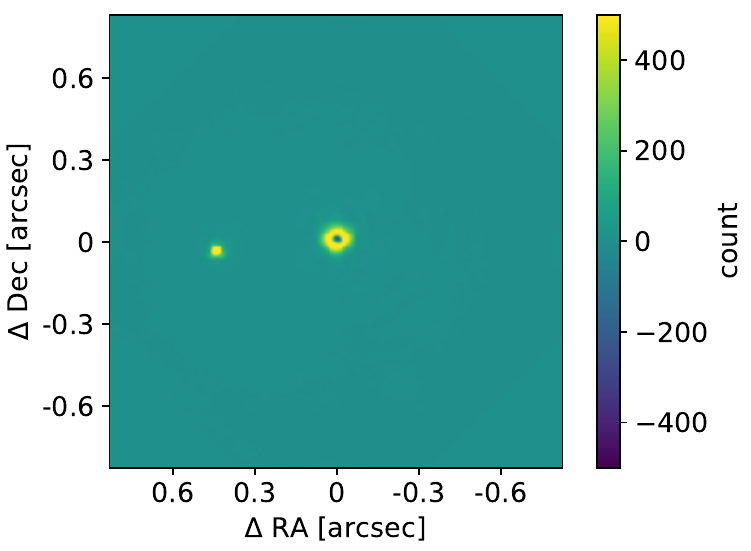}
\subcaption{Double-SDI result of omi Cet. A ring feature is visible around the central star, corresponding to the expected shock arising from the stellar pulsations.}
\label{fig: omi cet dSDI}
\end{minipage}
\end{tabular}
\caption[omi Cet result-1]{Flowchart of the SDI reduction for the omi Cet data.}
\end{figure}

\begin{figure}
\begin{tabular}{cc}
\begin{minipage}{0.5\hsize}
\centering
\includegraphics[width=0.95\textwidth]{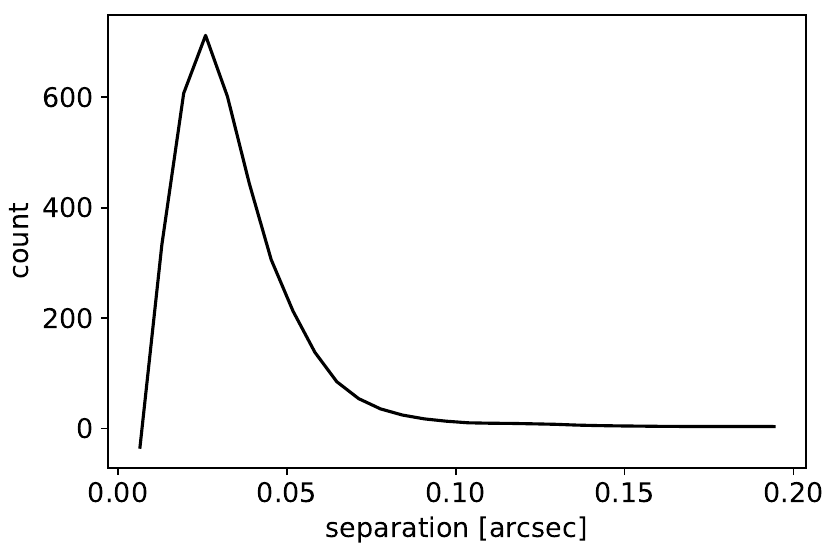}
\subcaption{Azimuthally-averaged radial profile of surface brightness around omi Cet after the double-SDI reduction (see Figure \ref{fig: omi cet dSDI}).}
\label{fig: omi cet radial profile}
\end{minipage}
\begin{minipage}{0.5\hsize}
\centering
\includegraphics[width=0.95\textwidth]{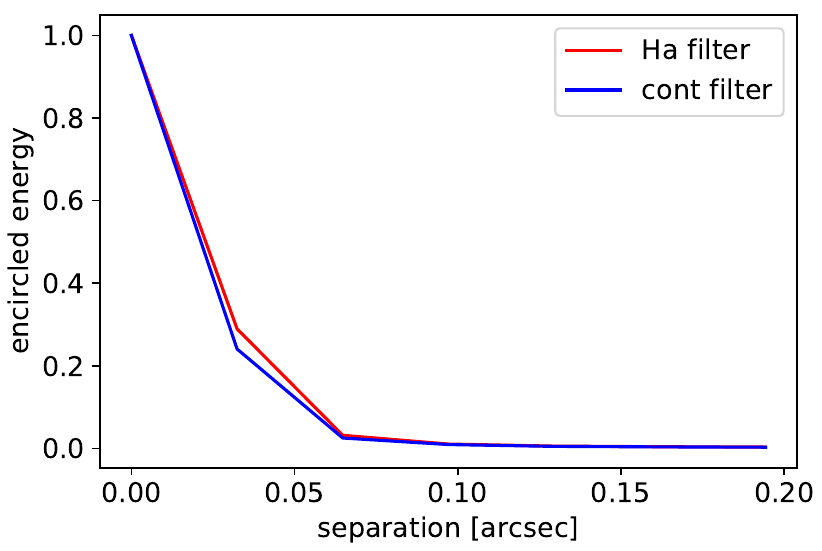}
\subcaption{Comparison of the encircled energies for combined H$\alpha$ (blue, Figure \ref{fig: omi cet comb_Ha}) and continuum (red, Figure \ref{fig: omi cet comb_cont}) images. The H$\alpha$ profile is brighter than that of the continuum profile at $\sim0.03^{\prime\prime}$.}
\label{fig: encircled energy omi cet}
\end{minipage}
\end{tabular}
\caption[omi Cet result-2]{PSF profiles of the omi Cet data.}
\end{figure}

Figure \ref{fig: omi cet raw} shows a single exposure raw H$\alpha$ image of omi Cet. We note that omi Cet B was detected in both combined H$\alpha$ and continuum images from one data cube (see Figures \ref{fig: omi cet comb_cont} and \ref{fig: omi cet comb_Ha}). Figures \ref{fig: omi cet SDI-state1} and \ref{fig: omi cet SDI-state2} compare the SDI-reduced images of State 1 and 2 respectively. 
Figure \ref{fig: omi cet dSDI} shows the double-SDI result and Figure \ref{fig: omi cet radial profile} plots an azimuthally-averaged radial profile of the double-SDI result. There is a ring-like feature at a separation of $\sim0.03^{\prime\prime}$. 
We also checked encircled energy of both combined H$\alpha$ and continuum images and H$\alpha$ profile is slightly brighter than the continuum profile at $\sim0.03^{\prime\prime}$ (see Figure \ref{fig: encircled energy omi cet}) , though the difference is marginal compared with the H$\alpha$ and continuum observations of $\eta$ Carinae \cite{Wu2017}. 
This feature represents either of the expected astrophysical feature - limb brightening of the shock feature at H$\alpha$ - or the difference of PSFs at wavelengths between the H$\alpha$ and the continuum filters.
To further test whether our data reduction can be affected by an artifact (e.g. different AO correction), we apply the double-SDI reduction to the SAO 105500 data (see Section \ref{sec: SAO 105500} for the result).
A scientific analysis of these resolved observations of the H$\alpha$ shock region will be presented in a forthcoming paper.

\begin{figure}
\begin{tabular}{cc}
\begin{minipage}{0.5\hsize}
\centering
\includegraphics[width=0.95\textwidth]{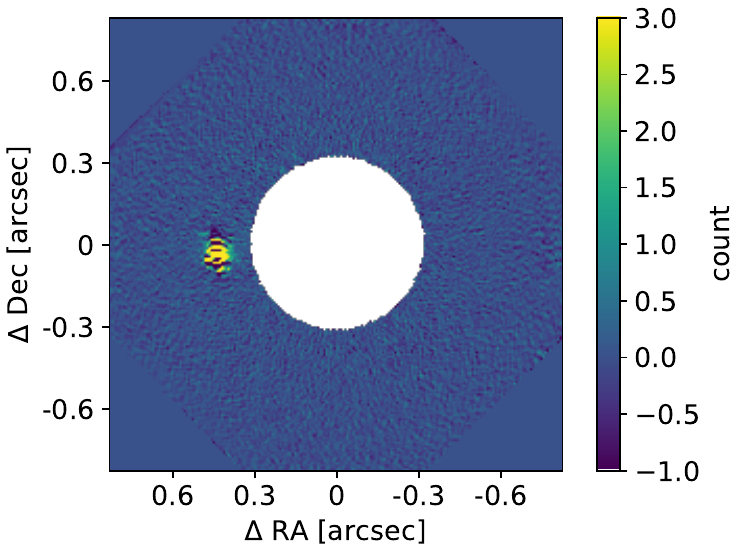}
\subcaption{ADI-reduced image (KL=3) after implementing double-SDI reduction (double-SDI+ADI). Central star is masked by the algorithm.}
\label{fig: omi cet SDI-ADI}
\end{minipage}
\begin{minipage}{0.5\hsize}
\centering
\includegraphics[width=0.95\textwidth]{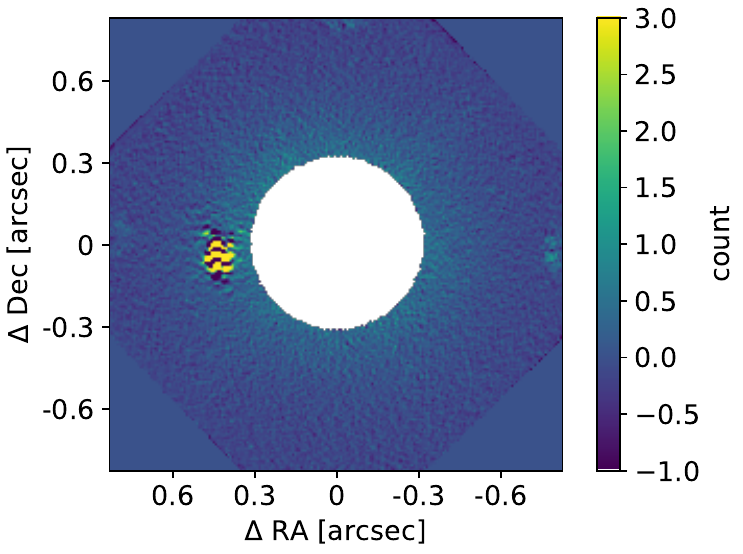}
\subcaption{Double-SDI-reduced image after implementing ADI (KL=3) reduction (ADI+double-SDI).}
\label{fig: omi cet ADI-SDI}
\end{minipage}
\end{tabular}
\caption[omi Cet result-3]{Flowchart of combining ADI and double-SDI for the omi Cet data.}
\end{figure}

Figures \ref{fig: omi cet SDI-ADI} and \ref{fig: omi cet ADI-SDI} compare the outputs of combining ADI and double-SDI reduction techniques. For ADI reduction we adopted KL=3 to show our outputs because we had a small number of data cubes and the larger KL number does not work properly. The smaller number of KL modes subtracts the starlight less efficiently and thus leaves a lot of residuals. Both of the reduction approaches resulted in almost the same outputs and detected omi Cet B with signal-to-noise ratios (SNRs) $>5000$ but with the ADI+double-SDI reduction (Figure \ref{fig: omi cet ADI-SDI}) there still remains some stellar halo at the inner working angle.
The noise was estimated by calculating the standard deviation within an annular region at different separations, which is used for signal-to-noise (SN) maps and detection limits.

\begin{figure}
    \centering
    \includegraphics[width=0.95\textwidth]{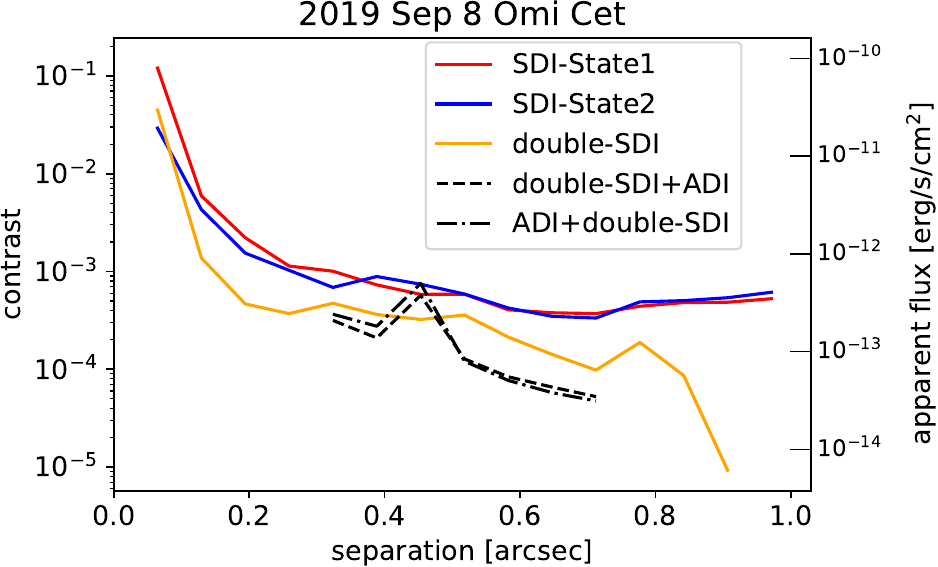}
    \caption[Contrast limits - omi Cet]{Comparison of 5$\sigma$ contrast limits of omi Cet data with a variety of reduction methods. `Double-SDI+ADI' and 'ADI+double-SDI' correspond to running ADI reduction after double-SDI reduction (Figure \ref{fig: omi cet SDI-ADI}) and the reverse order (Figure \ref{fig: omi cet ADI-SDI}), respectively.}
    \label{fig: omi cet contrast}
\end{figure}

Figure \ref{fig: omi cet contrast} compares 5$\sigma$ detection limits as a function of separation. The right vertical axis corresponds to apparent flux converted from contrast [erg/s/cm$^2$] assuming the $R$-band flux for omi Cet.
The humps at $0.4^{\prime\prime}-0.5^{\prime\prime}$ seen in the limits of double-SDI+ADI and ADI+double-SDI are affected by the existence of omi Cet B.
Regarding the throughput calculation to correct flux loss by the post-processing we assumed that a companion can be detected in only H$\alpha$ wavelength and that SDI reduction does not lose the signal of H$\alpha$. Therefore we injected fake sources in the H$\alpha$ images and calculated the throughputs of ADI reduction.
In case a companion is detected in both continuum and H$\alpha$ filters such as omi Cet B it is better to use photometric results from both combined H$\alpha$ and continuum images to accurately obtain H$\alpha$ intensity.

\subsubsection{RY Tau}

\begin{figure}
\begin{tabular}{cc}
\begin{minipage}{0.5\hsize}
\centering
\includegraphics[width=0.95\textwidth]{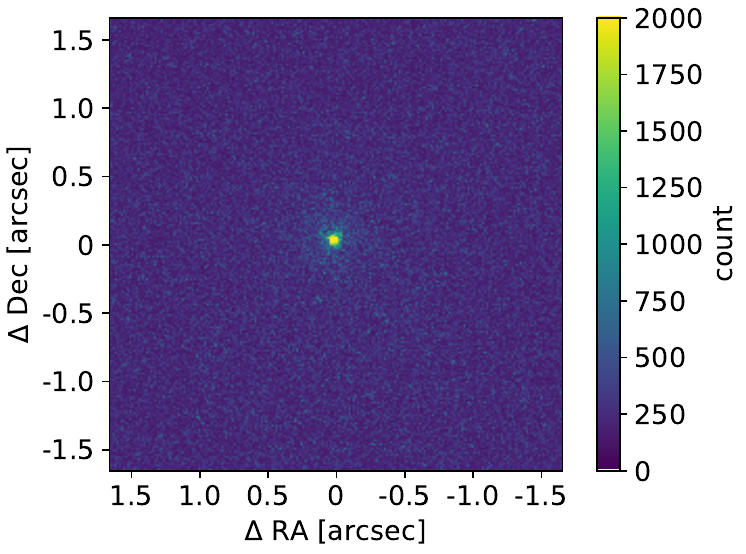}
\subcaption{Single H$\alpha$ image of RY Tau (exposure time = 1 sec) before the dark subtraction.}
\label{fig: RY Tau raw}
\end{minipage}
\begin{minipage}{0.5\hsize}
\centering
\includegraphics[width=0.95\textwidth]{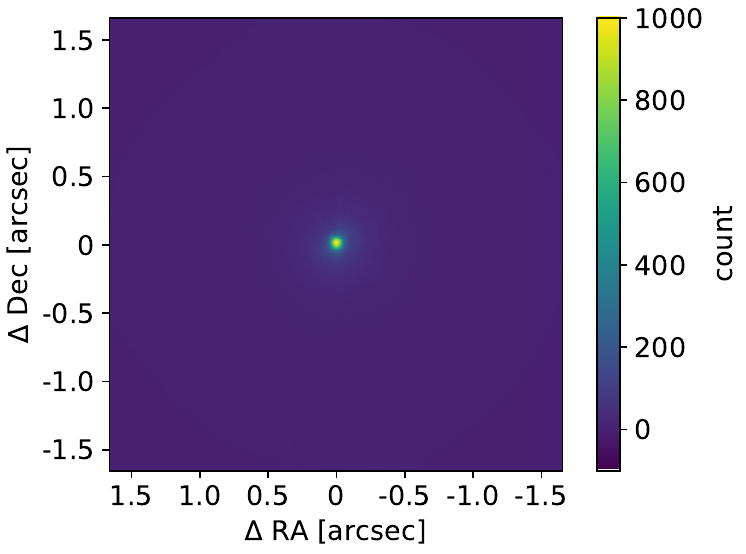}
\subcaption{SDI-reduced (H$\alpha$-continuum) image.}
\label{fig: RY Tau SDI-state1}
\end{minipage}
\end{tabular}
\caption[RY Tau result-1]{Flowchart of the SDI reduction for the RY Tau data.}
\end{figure}

\begin{figure}
\begin{tabular}{cc}
\begin{minipage}{0.5\hsize}
\centering
\includegraphics[width=0.95\textwidth]{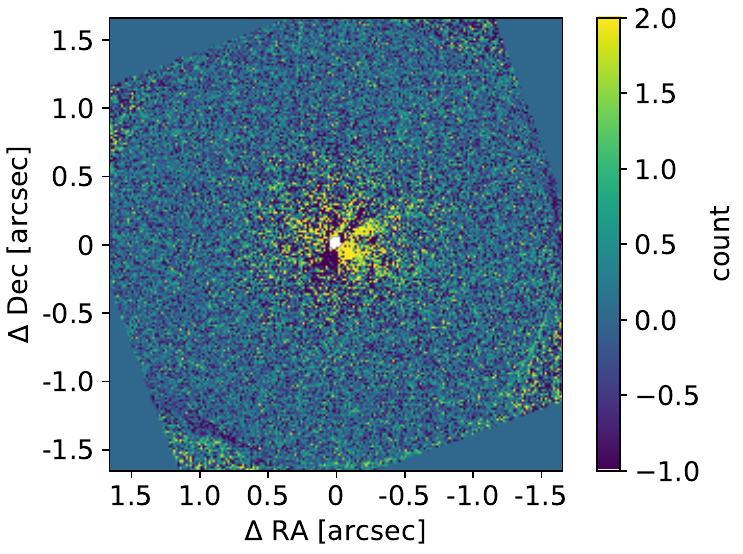}
\subcaption{ADI-reduced H$\alpha$ image.}
\label{fig: RY Tau ADI}
\end{minipage}
\begin{minipage}{0.5\hsize}
\centering
\includegraphics[width=0.95\textwidth]{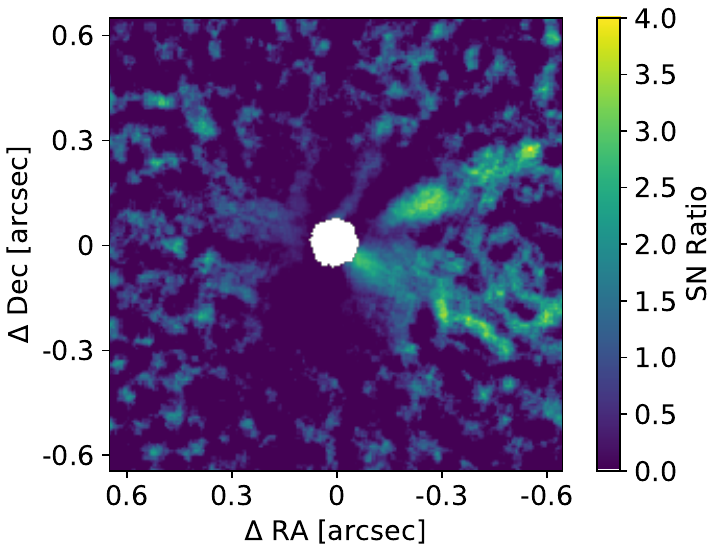}
\subcaption{Cropped SN map of Figure \ref{fig: RY Tau ADI}.}
\label{fig: snmap RY Tau ADI}
\end{minipage}\\
\begin{minipage}{0.5\hsize}
\centering
\includegraphics[width=0.95\textwidth]{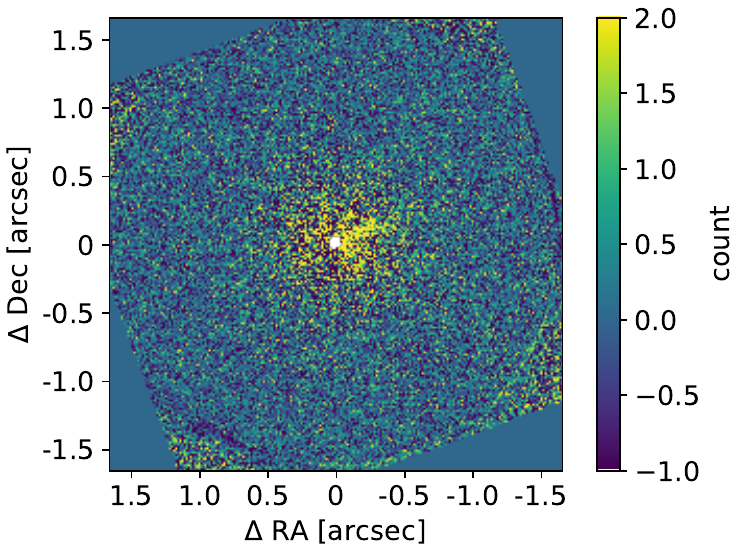}
\subcaption{ADI-reduced image after implementing SDI reduction.}
\label{fig: RY Tau SDI-ADI}
\end{minipage}
\begin{minipage}{0.5\hsize}
\centering
\includegraphics[width=0.95\textwidth]{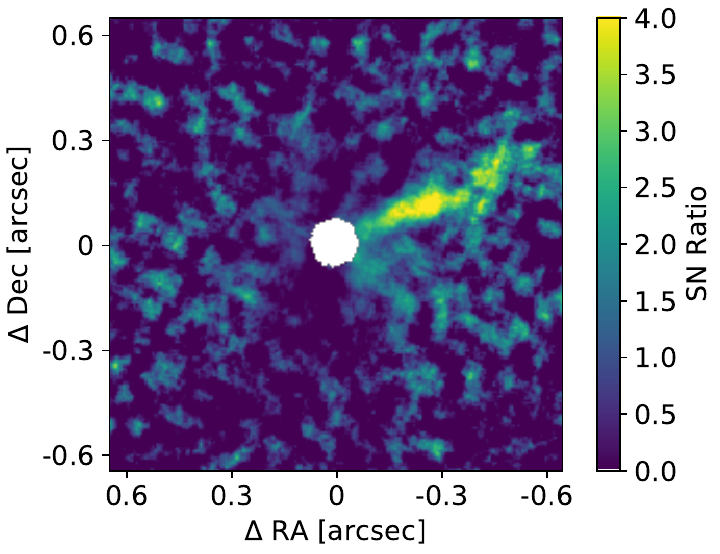}
\subcaption{Cropped SN map of Figure \ref{fig: RY Tau SDI-ADI}.}
\label{fig: snmap RY Tau SDI-ADI}
\end{minipage}\\
\begin{minipage}{0.5\hsize}
\centering
\includegraphics[width=0.95\textwidth]{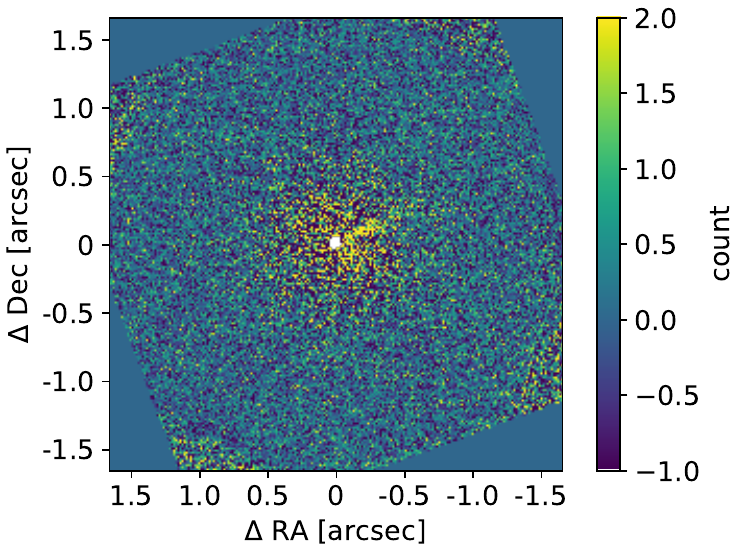}
\subcaption{SDI-reduced image after implementing ADI reduction.}
\label{fig: RY Tau ADI-SDI}
\end{minipage}
\begin{minipage}{0.5\hsize}
\centering
\includegraphics[width=0.95\textwidth]{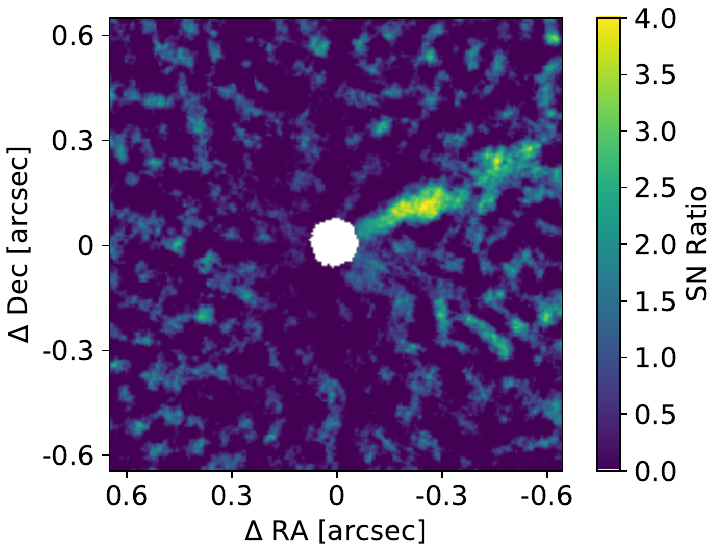}
\subcaption{Cropped SN map of Figure \ref{fig: RY Tau ADI-SDI}.}
\label{fig: snmap RY Tau ADI-SDI}
\end{minipage}
\end{tabular}
\caption[RY Tau result-2]{Flowchart of combining ADI and SDI for the RY Tau data.}
\end{figure}

Figure \ref{fig: RY Tau raw} shows a single exposure raw H$\alpha$ image of RY Tau.  Figures \ref{fig: RY Tau SDI-state1}, \ref{fig: RY Tau ADI} present a SDI-reduced and an ADI-reduced image of State 1. We note that RY Tau observation took only State 1 and we did not conduct the DDI reduction. 
Figures \ref{fig: RY Tau SDI-ADI} and \ref{fig: RY Tau ADI-SDI} compare the outputs of combining ADI and double-SDI reduction techniques. 
For ADI reduction we adopted KL=5 to show our outputs. Both of the reduction ways detected the jet with SNR$\sim$4-5 but its SNR is slightly higher in the SDI+ADI result (Figure \ref{fig: RY Tau SDI-ADI}). 
Figures \ref{fig: snmap RY Tau ADI}, \ref{fig: snmap RY Tau SDI-ADI}, and \ref{fig: snmap RY Tau ADI-SDI} present cropped SN maps of Figures \ref{fig: RY Tau ADI}, \ref{fig: RY Tau SDI-ADI}, and \ref{fig: RY Tau ADI-SDI}, respectively. Indeed a simple ADI reduction could detect the same feature to some extent as seen in the SDI+ADI image, but there is another feature that corresponds to speckles because of insufficient AO correction at optical wavelengths.

The most significant feature in our post-processed image extends $\sim0.3^{\prime\prime}$, which is consistent with the SPHERE/ZIMPOL H$\alpha$ observation \cite{Garufi2019}. 
The SPHERE observation also reported a fainter and more extended H$\alpha$ region extending $\sim1^{\prime\prime}$ and our SDI+ADI-reduced image marginally confirms the inner part ($\rho\sim0.3^{\prime\prime}-0.6^{\prime\prime}$) of this feature with SNR$\sim$3.

\begin{figure}
    \centering
    \includegraphics[width=0.95\textwidth]{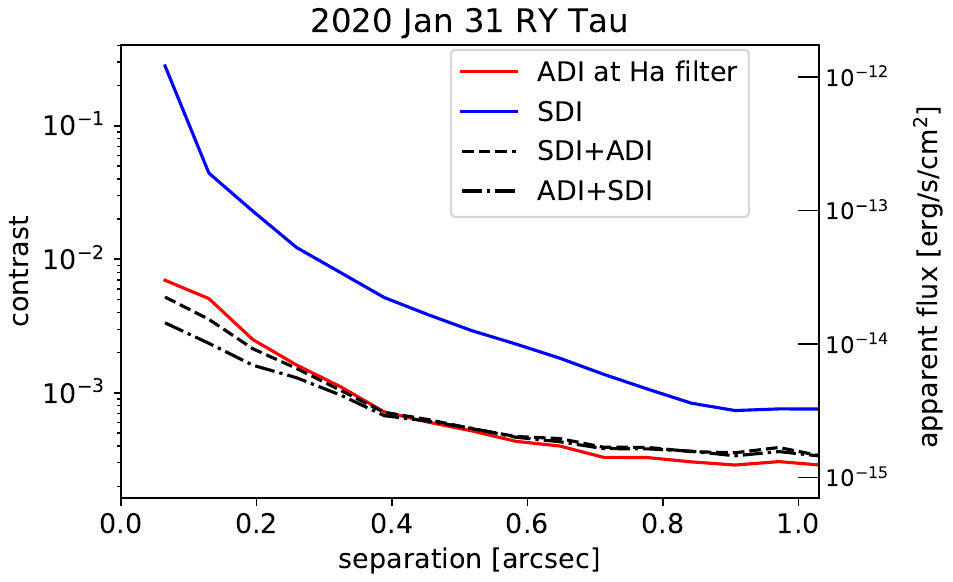}
    \caption[Contrast limits - RY Tau]{Comparison of 5$\sigma$ contrast limits of RY Tau data with a variety of reduction methods. `SDI+ADI' and 'ADI+SDI' correspond to running ADI reduction after SDI reduction (Figure \ref{fig: RY Tau SDI-ADI}) and the reverse order (Figure \ref{fig: RY Tau ADI-SDI}), respectively. }
    \label{fig: RY Tau contrast}
\end{figure}

Figure \ref{fig: RY Tau contrast} compares 5$\sigma$ detection limits as a function of separation. 
At larger separations ($\rho\gtrsim0.5^{\prime\prime}$) ADI reduction achieved higher contrast than the combination of ADI and SDI reduction because subtracting continuum image (camera 1) from H$\alpha$ image (camera 2) leaves the bias that increases the background noise. At inner separations ($\rho \lesssim 0.3$) the detection limit of SDI+ADI looks worse than that of ADI+SDI but this feature may be affected by higher SNR of the jet.
From a point of view of contrast the omi Cet observation (Figure \ref{fig: omi cet contrast}) achieved higher contrast than the RY Tau observation because of better wavefront sensing and AO correction, but RY Tau observation could achieve deeper detection limits in apparent flux. For future observations the beam switching with VAMPIRES and DDI reduction will help to achieve a better contrast level than the RY Tau result.
Compared with other H$\alpha$ high-contrast imaging observations, SCExAO+VAMPIRES has a similar sensitivity of H$\alpha$ to SPHERE/ZIMPOL \cite{Cugno2019, Zurlo2020}, MagAO \cite{Wagner2018}, and MUSE \cite{Haffert2019} and provides great opportunities for high-contrast H$\alpha$ imaging in the northern hemisphere.

\subsubsection{SAO 105500} \label{sec: SAO 105500}

\begin{figure}
\begin{tabular}{cc}
\begin{minipage}{0.5\hsize}
\centering
\includegraphics[width=0.95\textwidth]{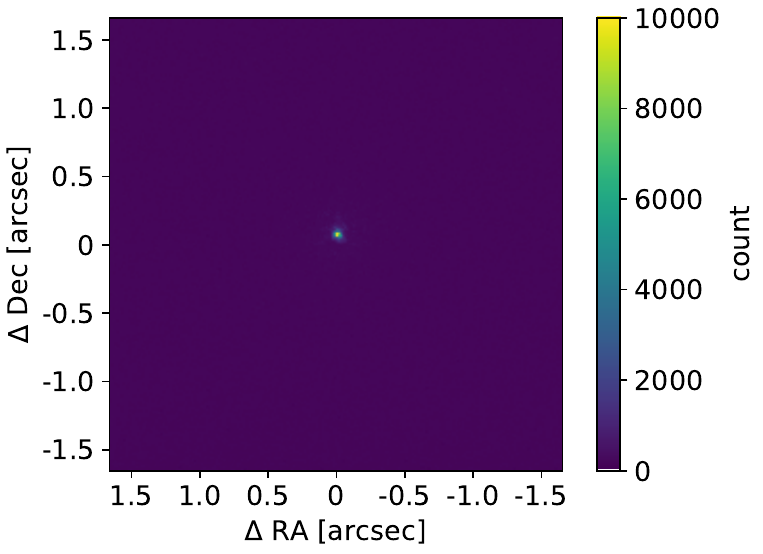}
\subcaption{Single H$\alpha$ image of SAO 105500 (exposure time = 20 msec) before the dark subtraction.}
\label{fig: sao 105500 raw}
\end{minipage}
\begin{minipage}{0.5\hsize}
\centering
\includegraphics[width=0.95\textwidth]{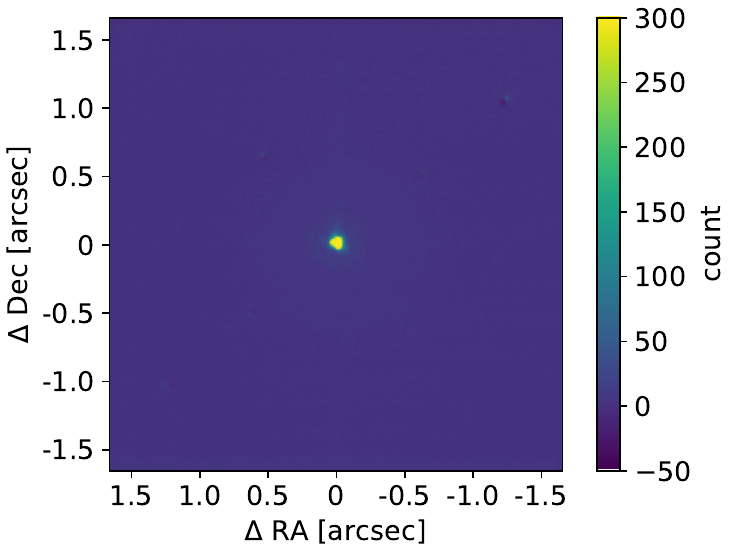}
\subcaption{SDI-reduced image of State1.}
\label{fig: sao 105500 state1}
\end{minipage}\\
\begin{minipage}{0.5\hsize}
\centering
\includegraphics[width=0.95\textwidth]{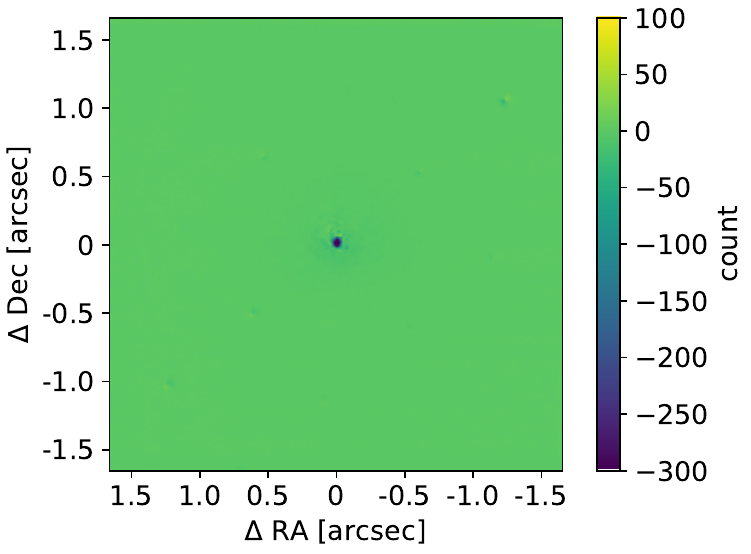}
\subcaption{SDI-reduced image of State2.}
\label{fig: sao 105500 state2}
\end{minipage}
\begin{minipage}{0.5\hsize}
\centering
\includegraphics[width=0.95\textwidth]{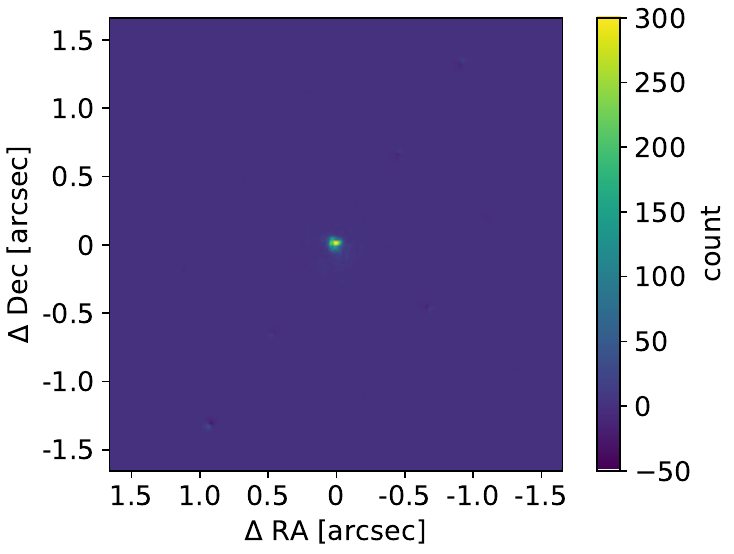}
\subcaption{Double-SDI-reduced image of SAO 105500.}
\label{fig: sao 105500 dsdi}
\end{minipage}\\
\begin{minipage}{0.5\hsize}
\centering
\includegraphics[width=0.95\textwidth]{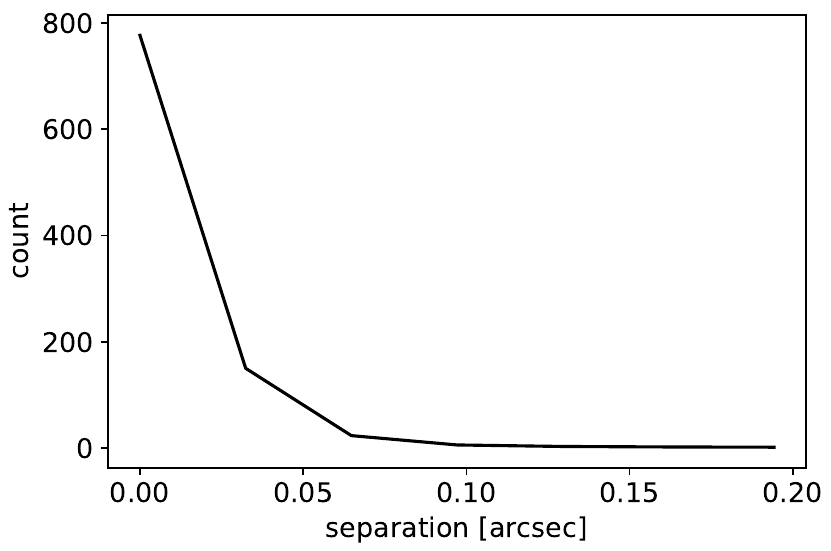}
\subcaption{Azimuthally-averaged radial profile of surface brightness around SAO 105500 after the double-SDI reduction (see Figure \ref{fig: sao 105500 dsdi}).}
\label{fig: sao 105500 radial profile}
\end{minipage}
\end{tabular}
\caption[SAO 105500 result]{Flowchart of the SDI reduction for the SAO 105500 data.}
\end{figure}

\begin{figure}
\begin{minipage}{0.5\hsize}
    \centering
    \includegraphics[width=0.95\textwidth]{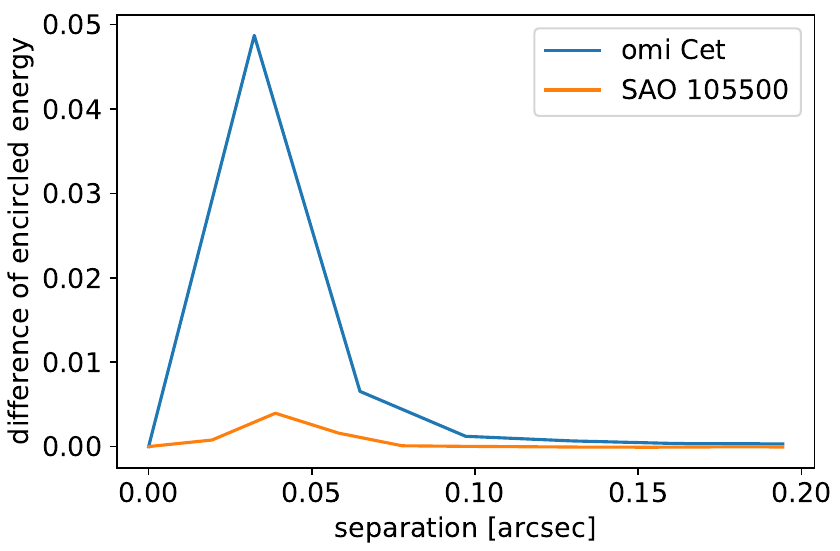}
    \caption[Comparison of encircled energy - omi Cet vs SAO 105500]{Difference of the encircled energies between H$\alpha$ and continuum PSFs.}
    \label{fig: difference encircled energy}
\end{minipage}
\begin{minipage}{0.5\hsize}
    \centering
    \includegraphics[width=0.95\textwidth]{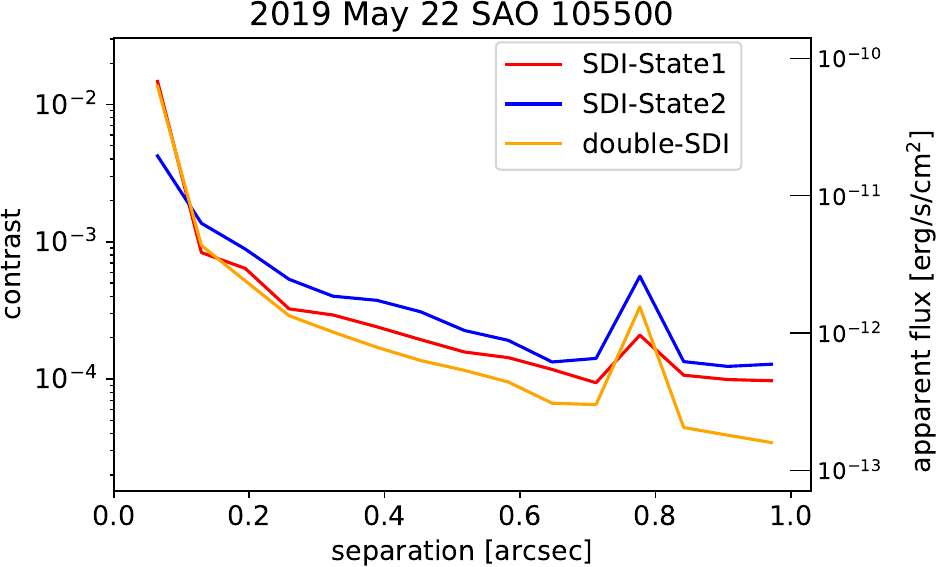}
    \caption[Contrast limits - SAO 105500]{Comparison of 5$\sigma$ contrast limits of SAO 105500 data with a variety of reduction methods.}
    \label{fig: sao 105500 contrast}
\end{minipage}
\end{figure}

For comparing the post-processing results of the other two targets, we reduced the SAO 105500 data. We note that because of small field rotation we could not conduct the ADI reduction and we show the SDI results in this section.
Figures \ref{fig: sao 105500 raw}, \ref{fig: sao 105500 state1}, \ref{fig: sao 105500 state2}, and \ref{fig: sao 105500 dsdi} show the single H$\alpha$ image and the SDI-State1, SDI-State2, and double-SDI results, respectively. 
We see the same tendency of the positive/negative counts at the central star at the SDI-reduced images in State 1/2 as the case of omi Cet (Figures \ref{fig: omi cet SDI-state1}, \ref{fig: omi cet SDI-state2}).
The residual at the center probably reflects the difference of its spectrum at wavelengths of the H$\alpha$-filter and the continuum-filter.
Figure \ref{fig: sao 105500 radial profile} shows an azimuthally-averaged radial profile of the surface brightness and there is no ring-like feature as seen in the case of omi Cet (Figure \ref{fig: omi cet dSDI}). 
Figure \ref{fig: difference encircled energy} compares difference of the encircled energies between H$\alpha$ and continuum PSFs. The slight rise in the SAO 105500 profile by $\sim0.4\%$ at $0.3^{\prime\prime}$ is likely caused by the difference of AO correction at H$\alpha$ and continuum wavelengths. The omi Cet profile has much greater difference by $\sim5\%$ at $0.3^{\prime\prime}$. Considering that AO correction to SAO 105500 works as effectively as to omi Cet, the difference profile at omi Cet may not only be affected by the difference of AO correction.
We need to model the expected H$\alpha$ shock around omi Cet to discuss in detail as mentioned in Section \ref{sec: Omi Cet}.

We also show the 5$\sigma$ detection limits of the SAO 105500 data in Figure \ref{fig: sao 105500 contrast}. The humps at $\sim0.8^{\prime\prime}$ are affected by diffraction patterns arising from the bright central star. 
These detection limits reach similar contrast levels to the omi Cet case because of almost the same AO efficiency.

\section{Summary and Future Prospects} \label{sec: Summary and Fugure Prospects}

We present high-contrast H$\alpha$ observations with Subaru/SCExAO+VAMPIRES. VAMPIRES adopts the beam-switching system where we can efficiently conduct SDI reduction by utilizing DDI technique. We also show the combination of ADI reduction with SDI with our engineering data sets of omi Cet and RY Tau.
Our double-SDI reduction could resolve a ring feature around omi Cet (corresponding to the expected shock arising from this star's pulsations) as well as detect omi Cet B in both continuum and H$\alpha$ filter, and SDI+ADI reduction could resolve a jet around RY Tau. We achieved $\sim10^{-4}$ at 0.5$^{\prime\prime}$ for omi Cet and $\sim5\times10^{-4}$ at 0.5$^{\prime\prime}$ for RY Tau, respectively.
Our detection limits in apparent flux are comparable to other high-contrast H$\alpha$ imaging with SPHERE/ZIMPOL, MUSE, and MagAO located in the southern hemisphere. Our instrument will provide great opportunities of implementing high-contrast H$\alpha$ explorations of northern targets.

In the next few years, AO188 will benefit from two major upgrades that will benefit SCExAO and VAMPIRES. AO188's 188-element deformable mirror will be replaced with a 64x64-element deformable mirror, which will improve significantly the wavefront correction at all wavelengths. The second upgrade is the addition of a NIR PyWFS inside AO188, which will be used to probe redder targets such as TTS \cite{Bond2018}. SCExAO is also being upgraded to perform PDI at NIR wavelengths, using a fast detector similarly to VAMPIRES, and also by doing spectro-polarimetry using CHARIS. A differential imaging mode is also envisioned to include a SDI mode at the Pa$\beta$ (1.28 $\mu$m) wavelength. In case of planet formation, for instance, theoretical models suggest that hydrogen emissions depend on several parameters such as number density of hydrogen, pre-shock velocity, extinction of the source, and filling factor of the emission \cite{Aoyama2019} and obtaining only H$\alpha$ cannot solve degeneracy between these parameters. Therefore Pa$\beta$ will help to solve the degeneracies and promote detailed discussions of accretion mechanisms.
Currently a limited number of instruments enable high-contrast imaging explorations of Pa$\beta$ (e.g. Keck/OSIRIS \cite{Uyama2017}) and SCExAO will be more useful instruments for hydrogen emission observations.
%\textcolor{red}{do we mention H$\alpha$+Pa$\beta$ simultaneous imaging?}
%SCExAO splits the incident light into optical and infrared bench and infrared rays are currently conveyed to CHARIS. 

\acknowledgments 
We would like to thank the anonymous referees for their constructive comments and suggestions to improve the quality of the paper.
The authors are grateful to Steven P. Bos for helping us calculate the Strehl ratios.
This research is based on data collected at the Subaru Telescope, which is operated by the National Astronomical Observatories of Japan.
We acknowledge with thanks the variable star observations from the AAVSO International Database contributed by observers worldwide and used in this research.
This research has made use of NASA's Astrophysics Data System Bibliographic Services.
This research has made use of the SIMBAD database, operated at CDS, Strasbourg, France.

TU acknowledges JSPS overseas research fellowship and JSPS overseas challenge program for young researchers.
The development of SCExAO was supported by JSPS (Grant-in-Aid for Research \#23340051, \#26220704 \& \#23103002), Astrobiology Center of NINS, Japan, the Mt Cuba Foundation, and the director's contingency fund at Subaru Telescope.

The authors wish to acknowledge the very significant cultural role and reverence that the summit of Mauna Kea has always had within the indigenous Hawaiian community. We are most fortunate to have the opportunity to conduct observations from this mountain.

\appendix
\section{Filter Information} \label{sec: Filter Information}
 
For high-contrast H$\alpha$ imaging with VAMPIRES we designed the bandpass of the filter to be able to transmit sufficient signal in light of various broadening effects. For example, the maximum accretion velocity of a Jupiter mass planet is $\sim 60$ km/s. To account for an object larger than a Jovian planet, we adopted a velocity of $\pm100$ km/s. In addition, the RV motion of a planet around such a star is $\lesssim10$ km/s and the RV motion of the star from the observer is $\lesssim50$ km/s.
Therefore, we set a budget of 200 km/s, which corresponds to a 0.8 nm wavelength range at 600 nm, which should cover all the astrophysical features one may like to study.

We found an off-the-shelf filter from Chroma that had specifications very close to our requirements (3 nm bandwidth; see also the manuscript webpage at \href{https://www.chroma.com/products/parts/h-alpha-3nm-bandpass}{Chroma}) and requested it be modified to have a FWHM bandwidth of 2 nm instead. This is still larger than the 0.8 nm needed and was chosen because of the uncertainty the manufacturer offered in the center wavelength of the filter and our ability to align it with respect to the beam. This filter was purchased and installed. However, the VLT/MUSE observation to study accretion from PDS 70 bc revealed that the effective width of H$\alpha$ is $\sim$100--120 km/s \cite{Haffert2019,Hashimoto2020}. We found that detecting such a signal with the 2~nm-bandwidth filter will require higher contrast, enough to detect the accreting planet. Therefore we decided to replace the previous H$\alpha$ filter with a new filter (1~nm-bandwidth) in January 2020. This bandwidth is the minimum acceptable value for the H$\alpha$ imaging with our assumptions.
Out of band, the optical density (OD) of the new H$\alpha$ filter is larger than 4 (OD4 means that the out-of-band suppression is 0.01\%) across 400-1200 nm, which covers the entire bandpass of the EMCCD. And similarly the suppression of the old H$\alpha$ and continuum filters was better than OD5 (the out-of-band suppression is 0.001\%) from 400-1000~nm.
The detailed information for the new H$\alpha$ filter is shown at \href{https://www.alluxa.com/optical-filter-catalog/ultra-narrow-bandpass/656-3-1-od4-ultra-narrow-bandpass.html}{Alluxa}.

\begin{figure}
    \centering
    \includegraphics[width=0.95\textwidth]{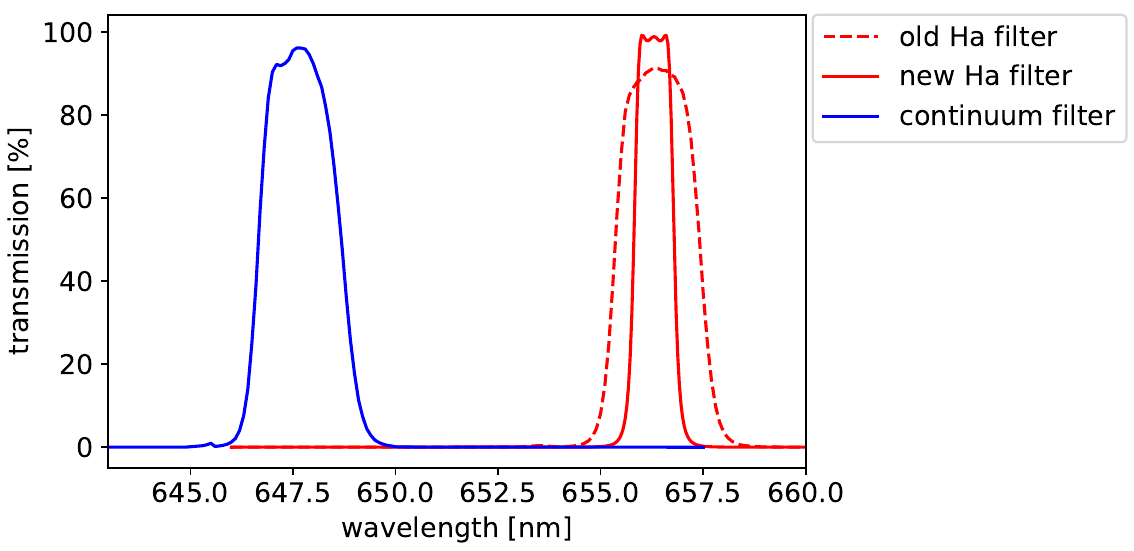}
    \caption[Transmission curves of the narrow-band filters]{Transmission curves of the narrow-band filters used in this study. The old H$\alpha$ filter was used for the SAO 105500 and omi Cet observation and the new H$\alpha$ filter was used for the RY Tau observation. The continuum filter has not been replaced.}
    \label{fig: sao 105500 contrast}
\end{figure}

%%%%% References %%%%%

\def\aj{AJ}%
         % Astronomical Journal
\def\actaa{Acta Astron.}%
         % Acta Astronomica
\def\araa{ARA\&A}%
         % Annual Review of Astron and Astrophys
\def\apj{ApJ}%
         % Astrophysical Journal
\def\apjl{ApJL}%
         % Astrophysical Journal, Letters
\def\apjs{ApJS}%
         % Astrophysical Journal, Supplement
\def\ao{Appl.~Opt.}%
         % Applied Optics
\def\apss{Ap\&SS}%
         % Astrophysics and Space Science
\def\aap{A\&A}%
         % Astronomy and Astrophysics
\def\aapr{A\&A~Rev.}%
         % Astronomy and Astrophysics Reviews
\def\aaps{A\&AS}%
         % Astronomy and Astrophysics, Supplement
\def\azh{AZh}%
         % Astronomicheskii Zhurnal
\def\baas{BAAS}%
         % Bulletin of the AAS
\def\bac{Bull. astr. Inst. Czechosl.}%
         % Bulletin of the Astronomical Institutes of Czechoslovakia 
\def\caa{Chinese Astron. Astrophys.}%
         % Chinese Astronomy and Astrophysics
\def\cjaa{Chinese J. Astron. Astrophys.}%
         % Chinese Journal of Astronomy and Astrophysics
\def\icarus{Icarus}%
         % Icarus
\def\jcap{J. Cosmology Astropart. Phys.}%
         % Journal of Cosmology and Astroparticle Physics
\def\jrasc{JRASC}%
         % Journal of the RAS of Canada
\def\mnras{MNRAS}%
         % Monthly Notices of the RAS
\def\memras{MmRAS}%
         % Memoirs of the RAS
\def\na{New A}%
         % New Astronomy
\def\nar{New A Rev.}%
         % New Astronomy Review
\def\pasa{PASA}%
         % Publications of the Astron. Soc. of Australia
\def\pra{Phys.~Rev.~A}%
         % Physical Review A: General Physics
\def\prb{Phys.~Rev.~B}%
         % Physical Review B: Solid State
\def\prc{Phys.~Rev.~C}%
         % Physical Review C
\def\prd{Phys.~Rev.~D}%
         % Physical Review D
\def\pre{Phys.~Rev.~E}%
         % Physical Review E
\def\prl{Phys.~Rev.~Lett.}%
         % Physical Review Letters
\def\pasp{PASP}%
         % Publications of the ASP
\def\pasj{PASJ}%
         % Publications of the ASJ
\def\qjras{QJRAS}%
         % Quarterly Journal of the RAS
\def\rmxaa{Rev. Mexicana Astron. Astrofis.}%
         % Revista Mexicana de Astronomia y Astrofisica
\def\skytel{S\&T}%
         % Sky and Telescope
\def\solphys{Sol.~Phys.}%
         % Solar Physics
\def\sovast{Soviet~Ast.}%
         % Soviet Astronomy
\def\ssr{Space~Sci.~Rev.}%
         % Space Science Reviews
\def\zap{ZAp}%
         % Zeitschrift fuer Astrophysik
\def\nat{Nature}%
         % Nature
\def\sci{Science}%
         % Science
\def\iaucirc{IAU~Circ.}%
         % IAU Cirulars
\def\aplett{Astrophys.~Lett.}%
         % Astrophysics Letters
\def\apspr{Astrophys.~Space~Phys.~Res.}%
         % Astrophysics Space Physics Research
\def\bain{Bull.~Astron.~Inst.~Netherlands}%
         % Bulletin Astronomical Institute of the Netherlands
\def\fcp{Fund.~Cosmic~Phys.}%
         % Fundamental Cosmic Physics
\def\gca{Geochim.~Cosmochim.~Acta}%
         % Geochimica Cosmochimica Acta
\def\grl{Geophys.~Res.~Lett.}%
         % Geophysics Research Letters
\def\jcp{J.~Chem.~Phys.}%
         % Journal of Chemical Physics
\def\jgr{J.~Geophys.~Res.}%
         % Journal of Geophysics Research
\def\jqsrt{J.~Quant.~Spec.~Radiat.~Transf.}%
         % Journal of Quantitative Spectroscopy and Radiative Transfer
\def\memsai{Mem.~Soc.~Astron.~Italiana}%
         % Mem. Societa Astronomica Italiana
\def\nphysa{Nucl.~Phys.~A}%
         % Nuclear Physics A
\def\physrep{Phys.~Rep.}%
         % Physics Reports
\def\physscr{Phys.~Scr}%
         % Physica Scripta
\def\planss{Planet.~Space~Sci.}%
         % Planetary Space Science
\def\procspie{Proc.~SPIE}%
         % Proceedings of the SPIE

\bibliography{library}   % bibliography data in report.bib
\bibliographystyle{spiejour}   % makes bibtex use spiejour.bst

\listoffigures
\listoftables

\end{spacing}

\end{document}